\documentclass[12pt]{iopart}

\usepackage{graphicx}

\begin{document}

\title[Vortices in Ionization Collisions by Positron Impact]
{Vortices in Ionization Collisions by Positron Impact}

\author{F. Navarrete, R. Della Picca, J. Fiol and R. O. Barrachina}
 \address{Centro At\'{o}mico Bariloche and Instituto Balseiro (Comisi\'{o}n Nacional de Energ\'{\i}a At\'{o}mica, CNEA, and Universidad Nacional de Cuyo, UNC), R8402AGP S. C. de Bariloche, R\'{\i}o Negro, Argentina}
 \eads{\mailto{barra@cab.cnea.gov.ar}}

\date{\today}
 
\begin{abstract}
The presence of vortices in the ionisation of hydrogen atoms by positrons at intermediate impact energies is investigated. The present findings show that a previously reported minima in the fully-differential cross section is the signature of a vortex in the continuum positron-electron-proton system. The behaviour of the real and imaginary parts of the complex-valued transition matrix is studied in order to determine and characterize the vortex in momentum space. The obtained information is translated to fully-differential ionisation cross sections, feasible of being measured with currently available techniques.
\end{abstract}


\section{Introduction}

In the last few years, the study and understanding of positron induced reactions with atoms and molecules have benefited from a new breed of experiments (Surko \etal 2005). The availability of intense mono-energetic positron beams and recent improvements in the measurement techniques are now routinely providing new evidence on such topics as Ps formation or positron annihilation and scattering. Besides the measurement of total and differential cross sections (e.g. Falke \etal 1997, Marler and Surko 2005, Laricchia \etal 2008), the future implementation of positron reaction microscopes would make it possible to obtain kinematically complete pictures of the product particles (Williams \etal 2010, Mueller \etal 2012). For instance, it might provide a way of analyzing such complex quantities as the fully differential cross section (FDCS) for positron-impact ionisation processes. It is noteworthy to point out that even in the simplest case of a hydrogenic target, the complete determination of the FDCS would require the measurement of four out of nine variables in the three-body momentum space, this latter number having been reduced by momentum and energy conservation and by the rotational symmetry about the collision axis.

Even before any positron reaction microscope becomes operational, experimental and theoretical studies of one and two-body spectra have already leaded to a clear understanding of some of the main features of the FDCS for positron-impact ionisation collisions. This is the case, for instance, of the electron capture to the continuum (ECC) effect (K\"{o}v\'{e}r and Laricchia 1998), whose characteristics in the momentum spectra of both the electron (Fiol and Barrachina 2011) and the recoil-ion (Barrachina and Fiol 2012) are now well understood.

However, up to our knowledge, the emergence of vortices in positron-impact ionisation collisions has not been researched neither experimentally nor theoretically. After being overlooked for decades, the presence of this kind of quantum structure in atomic and molecular processes was recently uncovered and even shown to be quite ubiquitous, being observed not only in electron-atom ionisation collisions (Macek \etal 2010, Colgan and Pindzola 2011), but also theoretically analyzed in the ionisation by the impact of ions (Macek \etal 2009, Ovchinnikov \etal 2011) and by intense electric pulses (Ovchinnikov \etal 2010). However, as far as we know, vortices have not been predicted or even searched for in positron-impact ionisation collisions.

It is noteworthy to mention that investigation of vortices in the case of electron impact presents additional complexities due to the indistinguishably of the projectile and target electrons. In the general case, the cross section may be written as the incoherent sum of the singlet and triplet terms. In the planar symmetric geometry studied in previous works one of these terms vanishes, and any vortex found in the other would produce a zero in the momentum distribution, as recognized earlier by Macek et al (2010). 
This is not longer true for the collinear case chosen in our article but, on the other hand, neither is the explicit choice of a particular geometry necessary with positrons since there is not exchange contribution.

More than two decades ago a deep minimum in the ionisation of atomic hydrogen by high-energy positrons was theoretically uncovered by Brauner and Briggs (1991). It was shown to appear when the electron and the positron emerged in the same direction at 45 degrees from the forward direction. In principle, it was attributed to the interference of two indistinguishable processes akin to the double binary-collision proposed by Thomas in 1927. In the mechanism proposed by Brauner and Briggs (1991), the positron and the electron leave a first binary collision with the same speed, only to end up moving in a collinear trajectory by the reorientation of any of them by an elastic scattering with the target nucleus. However, it is fair to mention that the minimum occurs at an electron speed smaller than that of the positron, an effect that increases for decreasing impact energies.

At that time this structure was not recognized as a zero of the cross section, even though the sudden drop of about three orders of magnitude seemed to be pointing towards a zero rather than just a deep minimum. On the light of similar effects which had been experimentally (Murray and Read 1993) and theoretically (Berakdar and Briggs 1994) observed in (e,2e) collisions, and which only quite recently were recognized as vortices (Macek \etal 2010), it can now be asserted with reasonable certainty that the effect theoretically discovered two decades ago by Brauner and Briggs (1991) in positron-impact ionisation collisions is a manifestation of a vortex.

Unfortunately, since this effect was observed at 10 and 100 keV impact energies, and shown to be absent at 250 eV, it only seemed to occur at very large energies, where the FDCS will be too small to be measurable in the coincidence conditions required by the collinear geometry (Brauner and Briggs 1991).

Thus, some few years ago, we addressed the question of whether there was another unforeseen minimum in the FDCS that would persist at moderate impact energies and therefore would be amenable to experimental observation (Della Picca \etal 2004). Actually, there is a deep minimum located at precisely the energy and emission angle where an electron could emerge by lying in the saddle-point of the positron and the residual target-ion potentials. Unmistakable signatures of this structure were also pinpointed in the positron dispersion and the recoiling of the target ion (Della Picca \etal 2005).

In the light of these evidences, the purpose of this article is to systematically investigate the precise location, structure and origin of this deep minimum and to demonstrate that it represents a fingerprint of a vortex in the transition matrix element (Navarrete \etal 2011).

\section{Electron - Positron - Ion Three-Body Kinematics}

We investigate the emission of a single electron from an atom of ionisation energy $\epsilon$ by the impact of a positron of initial velocity $\bi{v}$
Due to momentum conservation, the motion of the center-of-mass can be ignored and only two out of three positions or momenta are needed to completely describe the kinematics of the final three-body system. To this end, we will employ ion-centered coordinates, as defined by the relative positions $\bi{r}_\pm$ and momenta $\bi{k}_\pm$ of the electron (-) and the positron (+) with respect to the residual target of mass $M$. We will also use a set of Jacobi coordinates (Meyer 1999) given by the relative position $\bi{r}$ and momentum $\bi{k}$ of the electron with respect to the positron, and of their center-of-mass with respect to the residual target ion, namely $\bi{R}$ and $\bi{K}$. The relation between the ion-centered and the Jacobi coordinates and momenta is given by $\bi{r}_\pm = \mp \bi{r} / 2 + \bi{R}$ and $\bi{k}_\pm \approx \mp \bi{k} + \bi{K} / 2$, respectively. Here we are assuming that $M \gg 1$ (atomic units are used throughout this paper), so that $\bi{r}_\pm$ and $\bi{k}_\pm$ approximately represent the positions and momenta of the electron and the positron in a Laboratory reference frame. The fully differential cross section reads
\begin{equation}
\frac {\mathrm{d} \sigma} {\mathrm{d} \bi{k_-} \, \mathrm{d} \bi{k_+}} = \frac {(2 \pi)^4} {v} \left| T_{if} \right|^2 \delta \left(\frac {K^2} {4} + (k^2 + \epsilon) - \frac {v^2} {2}  \right) \; ,
\label{FDCS}
\end{equation}
where $T_{if} = \langle \Psi^-_f | V_i | \Psi_i \rangle$ is the transition matrix element. Here, the initial state $\Psi_i$ accounts for the bound state of the target and the free motion of the positron with respect to the target. $V_i$ is the interaction of the projectile with both particles in the target, namely $V_i = Z/{\rm r}_+ - 1/{\rm r}$, where $Z$ is the effective charge of the receding ion. Let us recall that the study of the three-body system with Coulomb interactions is one of the basic unsolved problems of Quantum Mechanics. Thus some kind of approximation is required for the description of the final continuum state $\Psi^-_f$. Here we choose to describe this final state by means of a correlated C3 wavefunction (Garibotti and Miraglia 1980, Brauner and Briggs 1986),
\begin{equation}
\Psi_{C3} = \psi_{-Z} \left(\bi{k}_+, \bi{r}_+ \right) \times \psi_{Z} \left( \bi{k}_-, \bi{r}_- \right) \times \frac { \psi_{1/2} \left(\bi{k}, \bi{r} \right)} {\psi_{0} \left(\bi{k}, \bi{r} \right)}
\; , \label{C3}
\end{equation}
where
\[
\psi_\mu \left( \bi{p}, \bi{q} \right) = (2 \pi)^{-3/2} N( \mu / {\rm p} ) \,_1F_1 \left( - i \mu / {\rm p} , 1 ; - i \left( {\rm p} {\rm q} + \bi{p} \cdot \bi{q} \right) \right) e^{i \bi{p} \cdot \bi{q}} \; ,
\]
with normalization $N(\nu) = \Gamma (1 + i \nu) \exp ( \pi \nu /2 )$, is the continuum state for a two-body system of relative position $\bi{q}$, momentum $\bi{p}$ and unity reduced mass, interacting via a Coulomb potential $V({\rm q}) = - \mu / {\rm q}$.

Since the use of a C3 wavefunction is based on a perturbative approach it provides a better description of the ionisation process at large incident positron velocities. However, as explained in the introduction, the cross section magnitudes decrease strongly as the incident velocities increase, making very difficult any measurement of the FDCS. As a compromise, in this situation we will perform calculations at two collision energies: 100 and 200~eV. While it is expected that the present theory would provide more accurate results at the larger energy, it is more likely that experimental confirmation will be reported at the lower energy of 100~eV.

This approximation was already employed by us to successfully describe the electron capture to the continuum (ECC) effect in this same single ionisation collision by positron impact (Fiol \etal 2001, Della Picca \etal 2006). Similar C3 approximations provided the first theoretical evidence of a deep minimum in positron-impact ionisation collisions at high energies (Brauner and Briggs 1991) and, with modified Sommerfeld $\nu$ parameters, in (e,2e) processes (Berakdar and Briggs 1994). It also provided our first uncovering of a deep minimum of the FDCS at a saddle-point of the electron-positron-ion system in the ionisation of atoms by the impact of positrons of moderate energies (Della Picca \etal 2004, 2005, 2006). Also, a variation of the C3 wavefunction, based on the works of Alt and Mukhamedzhanov (1993), modified to verify the correct asymptotic conditions has been employed to address (e,2e) processes (Berakdar and Briggs, 1994, 1994a), and very recently employed by Macek \etal (2010) to investigate the appearance of vortices in electron-atom collisions in a symmetric geometry. It is noteworthy that this modification reduces to the C3 wavefunction (eq.~\ref{C3}) for the collinear geometry employed in this work. 

\section{Collinear Geometry}

Note that by energy conservation, the number of relevant scalar variables of the FDCS in Eq.~\eref{FDCS} is reduced from six to five. Furthermore, in the absence of polarized targets or projectiles, the collision is axisymmetrical about the initial velocity $\bi{v}$; and the dimension of the problem is further reduced to four. But even the mere representation of such a multidimensional object would be a formidable task, and this is so at two very basic levels. On one hand, there is the problem of choosing a set of four variables that would better represent the problem under study. Naturally, there is a large degree of freedom in the selection of these quantities. Some of these choices are related to different types of experiments. This is the case of the so-called Rochester geometry (Baier \etal 1969) where the cross section is written in terms of the polar angles of two outgoing particles, their relative azimuthal angle and the energy of one of them. This choice is complete in the sense that any other set of independent variables can be related to this one.

The second difficulty with the FDCS is that it can not be represented in a 3-D space. The search for a practical solution to this conundrum is as old as Science and Art themselves (Malcolm 2004, Pauwels 2006). One way of making this representation manageable, but not the only option, is to reduce the dimension of the cross section. And there are two different ways of accomplishing this task. One option is to integrate the cross section over two or more of its variables. This is the case, for instance, in any single particle spectroscopy technique, where only the energy and/or the emission angle of one of the particles in the final state is investigated. The second option, and the one that we will adopt in this article, is to fix one or more variables to a particular value or condition. A very familiar and convenient choice when analyzing processes with two similar particles in the final state, as in $(e, 2e)$ and $(p, 2p)$ collisions (see, for instance, Lim and McCarthy 1966), is the so-called ``symmetric geometry'' introduced by Gottschalk \etal in 1965 for the study of proton-proton bremsstrahlung (Gottschalk \etal 1965). In this geometry, the two outgoing identical particles are forced to have equal energies and polar angles. In our case, when applied to the electron and the positron in the final state, the use of the Gottschalk symmetry would mean that $\bi{k}$ is forced to be perpendicular to the plane formed by $\bi{v}$ and $\bi{K}$. This geometry is clearly equivalent to intersecting the FDCS with a 2D surface, so as to make it dependent on only two parameters.

Another standard geometry that accomplishes the same reduction is the ``energy sharing'' or collinear arrangement, where the electron and the positron are forced to move along the same direction, {\it i.e.} $\bi{k}_- \cdot \bi{k}_+ = k_- k_+$. Contrary to Gottschalk's Symmetric Geometry, here $\bi{k}$ is parallel (or antiparallel) to $\bi{K}$. Note, that the variation of $\bi{k}_-$ and $\bi{k}_+$ is not independent, since they have to share the energy according to $k_-^2 + k_+^2 \approx v^2 - 2 \epsilon$. In this paper we will employ this latter geometry.

\section{Vortices in positron-atom ionisation collisions}

Vortices are ubiquitous in classical and quantum systems with large numbers of particles. Here, however, we are interested in a different kind of vortices that occur in the wave function of a few-body quantum system when considered in the framework of de Broglie - Bohm theory (Bohm 1952, D\"{u}rr and Teufel 2009) or Madelungs's Hydrodynamic interpretation (Madelung 1926, Ghosh and Deb 1982). In 1926 Madelung demonstrated that the time evolution of a wave function $\Psi$, as given by the Schr\"{o}dinger equation, is equivalent to that of a perfect fluid of density $\rho = |\Psi|^2$ and velocity $\bi{u} = {\rm Im} \left( \nabla \ln \Psi \right)$ described by Euler equations with a pressure tensor
\[
P_{ij} = - \frac {\rho} {4m} \; \nabla \otimes \nabla \ln \rho
\]
The flow is shown to be irrotational except at vortex singularities where the velocity diverges.

In the single ionisation of an atom by positron impact, vortices might be formed in the three-body time-dependent wave function $\Psi (\bi{r}_+ , \bi{r}_- , t)$ at the coalescence region i.e. when all the intervening particles are close together. They should emerge either as a closed submanifold or in pairs of opposite circulation (Bialynicki-Birula \etal 2000) and collapse at later times, but some of them can eventually survive up to the asymptotic regime at macroscopic distances (Macek 2009). In this case, according to the imaging theorem (Dollard 1971, Macek \etal 2010)
\[
|T_{if}| \propto \lim_{t\rightarrow \infty} t^3 \, | \Psi (\bi{k}_+ t, \bi{k}_- t, t) |
\]
they will directly manifest themselves as zeros of the ionisation matrix element $T_{if}$.

\begin{figure}
\centering{}\includegraphics[width=.42\linewidth]{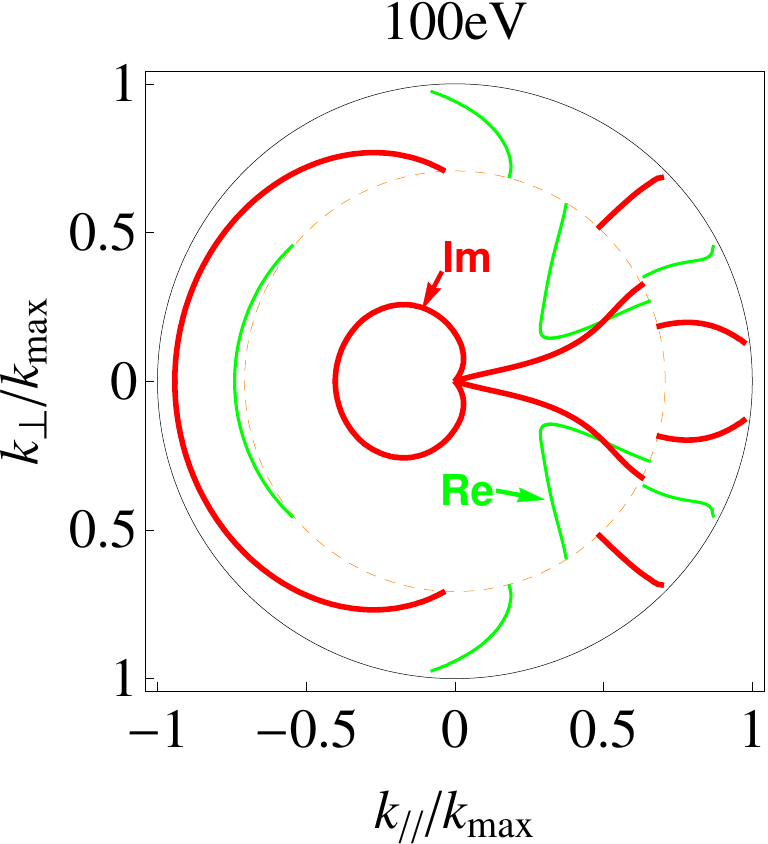} \qquad\includegraphics[width=.42\linewidth]{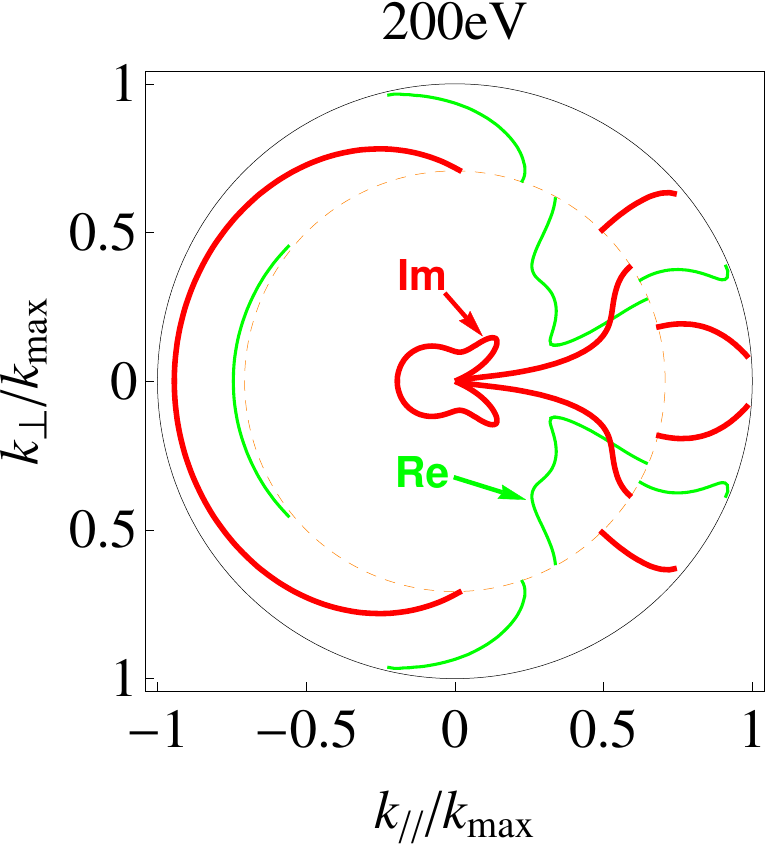}
\caption{(color online) Lines of zero real and imaginary parts of the reduced transition matrix element $\tilde{T}_{if}$ (see text) in a collinear geometry for the ionisation of an Hydrogen atom by the impact of a positron of 100 (Left) and 200~eV (Right). $k_{||}$ and $k_\perp$ are the components of the electron momentum $\bi{k}_e$ parallel and perpendicular to the initial velocity of the positron, respectively. The axis are normalized to the incident momentum $k_{\mathrm{max}}$.\label{real_imaginary}}
\end{figure}

Since both the real and imaginary parts of $T_{if}$ have to vanish at a vortex, such structure will eventually span a submanifold with codimension 2. By employing a collinear geometry, we are reducing the number of independent scalar variables from four to two, and therefore vortices should have to show up as isolated zeros. With this in mind, we conduct a systematic search of vortices in positron-impact ionisation collisions at intermediate energies, by looking for crossings of the projection onto the collinear geometry of the submanifolds where the real and the imaginary parts of $T_{if}$ vanish. Within this geometry, the real and imaginary parts of the normalization factors $N(Z / k_-)$ and $N(1 / 2k)$ vanish on an infinite number of circles accumulating at $k_- = 0$ and $k_- = k_{+} $. These conditions correspond to the well-known mechanisms of electron excitation to the continuum (EEC) and electron-capture-to-the-continuum (ECC), respectively. This infinite number of non-crossing lines might mask the emergence of a vortex. Thus, in figure \ref{real_imaginary} we choose to show only the lines where the real and imaginary parts of the reduced transition matrix element $\tilde{T}_{if} = T_{if} / N(Z / k_-) N(1 / 2k)$ vanishes. 
Obviously, these lines do not coincide with those corresponding to $T_{if}$, but their crossings do. Thus, their use is instrumental for our task of locating single zeros of the ionisation matrix element. As it is shown in figure \ref{real_imaginary}, in the case of the ionisation of atomic Hydrogen by positrons of 100 and 200~eV, this only occurs at a single point, when the electron is in the close vicinity of the saddle point potential of the positron - ion continuum system. 

\begin{figure}
\centering{}
\includegraphics[width=.495\linewidth]{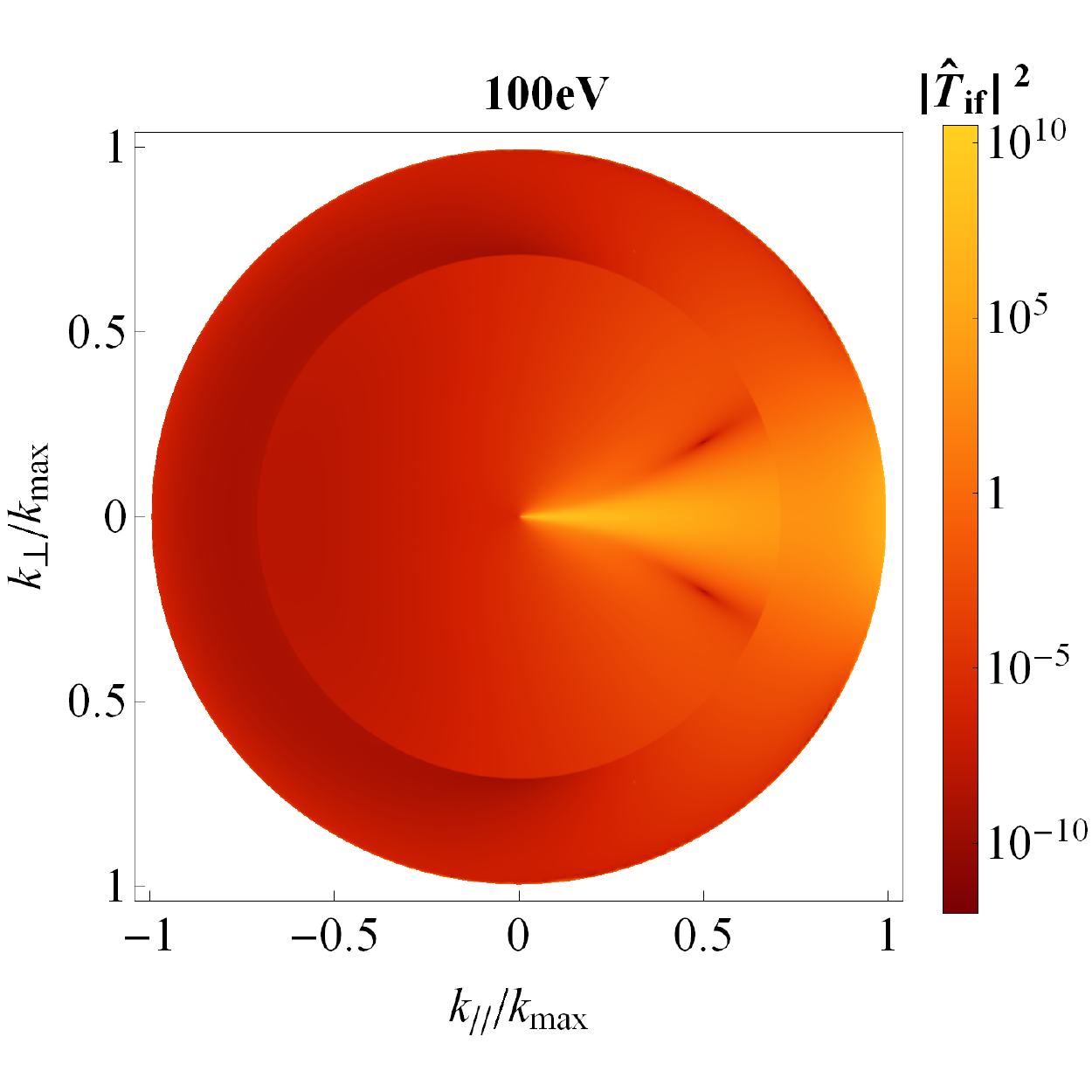}\includegraphics[width=.495\linewidth]{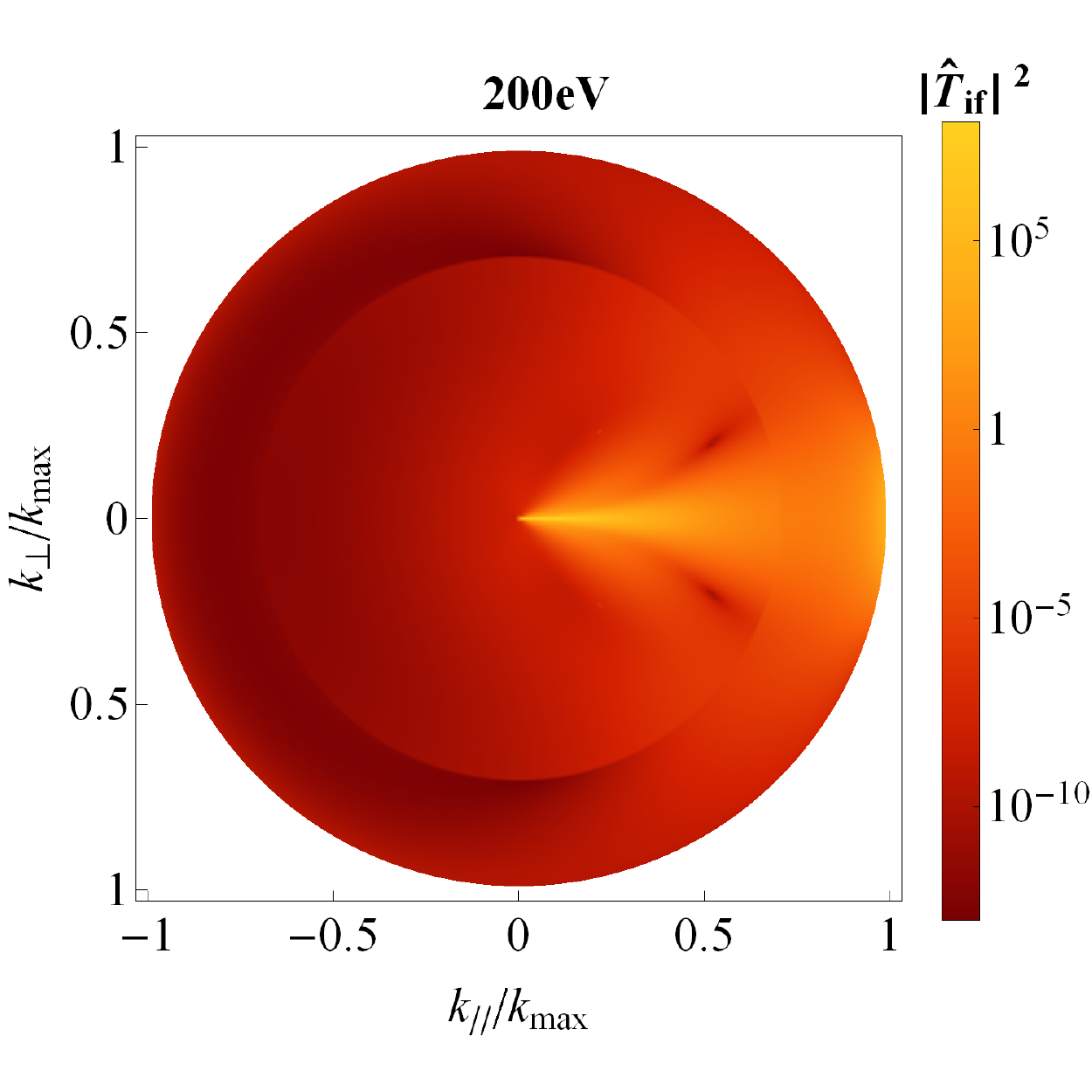}
\caption{(color online) Square modulus of the amplitude of the reduced transition matrix element $|\tilde{T}_{if}|^{2}$ (see text) for the ionisation of an Hydrogen atom by the impact of a positron of 100 (Left) and 200~eV (Right). Conditions are set to the collinear geometry configuration discussed above. $k_{||}$ and $k_\perp$ are the components of the electron momentum $\bi{k}_e$ parallel and perpendicular to the initial velocity of the positron, respectively.}
\label{amplit_tif}
\end{figure}

In \fref{real_imaginary}, at the point where the zeros of the real and imaginary part of the reduced transition-matrix cross each other, the amplitude vanish. This is clearly observed in figure \ref{amplit_tif}, confirming our expectations that the deep minimum previously found by us at 100~eV (Della Picca \etal 2005) actually represents a zero of the FDCS.
Moreover, the present results confirm that this is not an isolated effect at 100~eV but that the same features are obtained at larger energies. Although some differences in the FDCS are observed for the two energies investigated they, present a similar qualitative behavior, showing vortices at about the same electron emission direction.

\begin{figure}
\centering{}\includegraphics[width=.45\linewidth]{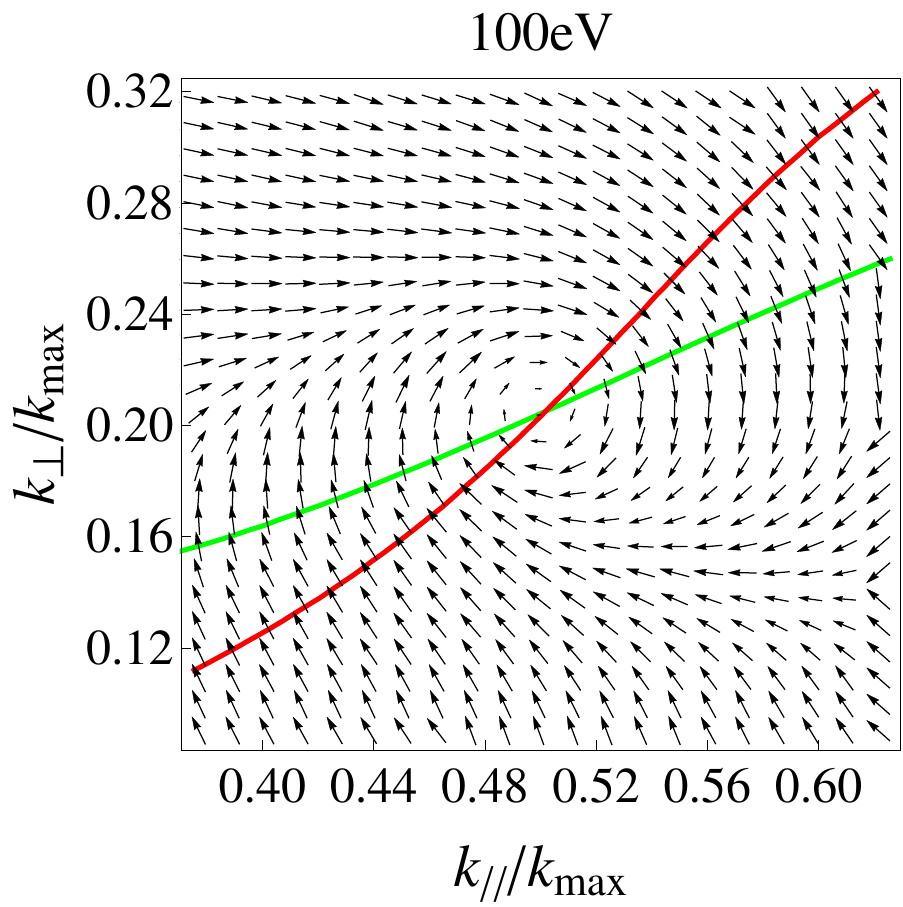}
\caption{Generalized velocity field of the reduced transition matrix element $\tilde{T}_{if}$ (see text) in a collinear geometry for the ionisation of an Hydrogen atom by the impact of a positron of 100~eV.}
\label{velocity_field}
\end{figure}

The generalized velocity field
\[
\bi{u} = \mathrm{Im} \frac {\nabla_{\bi{k}_+ , \bi{k}_-} T_{if}} {T_{if}}
\]
can be shown to be normal to the submanifolds where the real or the imaginary parts of the wave-function vanishes. Thus it produces a circulation along any closed contour encircling the vortex that must be quantized to a multiple of $2 \pi$ in order to assure the single valuation of the wave-function (Bialynicki-Birula \etal 2000). The velocity field associated to the reduced ionisation matrix element $\tilde{T}_{if}$ is shown in figure \ref{velocity_field}. In fact, the numerical integration of the generalized velocity field around the vortex yielded a circulation equal to $2 \pi$.

Further calculations to be shown in a future more comprehensive article let us confirm the appearance of other vortices at higher energies, as the one reported by Brauner and Briggs (1991). But only the one discussed here remains at intermediate and low energies, being visible at energies as low as 30 eV. Furthermore, its location in momentum space does not significantly depend on the impact energy.

\section{Conclusions}

In this article we have uncovered the presence of a vortex in positron-impact ionisation collisions at intermediate energies that could be observed with reaction-microscope or coincident electron-positron spectroscopy techniques. Vortices are ubiquitous in classical and quantum systems with large (macroscopic) number of particles. They are routinely observed in connection with superfluidity, superconductivity, or Bose-Einstein condensation, and are described by introducing nonlinear terms in the Hamiltonian. 
The main qualitative differences of these examples with the vortex investigated in this work are, on one hand the extreme simplicity of the present three-body system,  consisting of an electron, a positron, and a proton. Additionally, the 
time-evolution is governed exactly by the Schr\"{o}dinger equation with Coulomb interactions, with no ad-hoc potential models or nonlinear terms.
In fact, our theoretical prediction might prompt the experimental search of positron-induced vortices, not only providing a possible confirmation of our findings, but also motivating the study of the topology and dynamics of these intriguing quantum structures. 

The vortex investigated here is located in a configuration where the electron sits in the vicinity of the positron - ion potential saddle point.
In this sense, it is akin to a deep minima theoretically observed by Brauner and Briggs (1991) at 45 degrees in a similar collinear geometry with significantly larger impact energies of 10 and 100 keV. In both cases, vortices seem to be located at significant points related to particular collisions mechanisms, as the saddle-point and Thomas processes, respectively. Whether this represents a rule or a fortuitous coincidence remains to be investigated.

Finally, let us point out that all the available studies of vortices in ionisation processes, and the one undertaken here is not an exception, have relied on some kind of geometry that projects the FDCS onto a two-dimensional plane. This means that the isolated zeros observed so far in electron and positron-impact ionisation collisions are only the cut of minimum dimensionality of more complex structure spanning a submanifold of codimension 2 in the 4D space of independent scalar variables of the FDCS. The future research of this complex structure would clearly represent a formidable but inescapable task in order to achieve a fundamental understanding of these few-body fascinating structures.

\section*{Acknowledgments}

This work was supported by the Consejo Nacional de Investigaciones Cient\'{\i}ficas y T\'{e}cnicas (Grant no PIP 112-200801011269) and Universidad Nacional de Cuyo (grant 06/C348). The authors are also members of the Consejo Nacional de Investigaciones Cient\'{\i}ficas y T\'{e}cnicas (CONICET), Argentina.

\References
\item[] Alt, E O and  Mukhamedzhanov A M 1993 \textit{Phys. Rev.}~A \textbf{47} 2004-2022
\item[] Baier R, K\"{u}hnelt H and Urban P 1969 \NP {\bf B11} 675
\item[] Barrachina R O and Fiol J 2012 \jpb {\bf 45} 065202
\item[] Berakdar J and Briggs J S 1994 \jpb {\bf 27} 4271
\item[] Berakdar J and Briggs J S 1994a \PRL {\bf 72} 3799
\item[] Bialynicki-Birula I, Bialynicka-Birula Z \'Sliwa C 2000 \PR A {\bf 61} 032110
\item[] Bohm D 1952 \PR {\bf 85} 166, 180
\item[] Brauner M and Briggs J S 1986 \JPB {\bf 19} L325
\item[] Brauner M and Briggs J S 1991 \JPB {\bf 24} 2227
\item[] Bruns H 1895 {\it Das Eikonal}, Berichte \"{u}ber die Verhandlungen der k\"{o}niglich s\"{a}chsischen Gesellschaft der Wissenschaften zu Leipzig {\bf 21} 323
\item[] Colgan J and Pindzola M S 2011 {\it J. Phys.: Conf. Ser.} {\bf 288} 012001
\item[] Della Picca R, Fiol J and Barrachina R O 2004 {\it Evidence of saddle-point electron emission in an ionization process induced by positron impact}, 8th Workshop on Fast Ion - Atom Collisions (Debrecen, Hungary)
\item[] Della Picca R, Fiol J and Barrachina R O 2005 \NIM {\it B} {\bf 233} 270
\item[] Della Picca R, Fiol J and Barrachina R O 2006 \NIM {\it B} {\bf 247} 52
\item[] Dollard J D 1971 Rocky Mountain J.~Math. 1, 5-88
\item[] D\"{u}rr D and Teufel S 2009 {\it Bohmian Mechanics} (Berlin: Springer)
\item[] Falke T, Brandt T, Kuhl O, Raith W and Weber M 1997 \jpb {\bf 30} 3247
\item[] Fiol J, Rodr\'{\i}guez V D and Barrachina R O 2001, \jpb \textbf{34}, 933-944
\item[] Fiol J and Barrachina R O 2011 \jpb {\bf 44} 075205
\item[] Garibotti C R and Miraglia J E 1980 \PR A 21 572
\item[] Ghosh S K and Deb B M 1982 {\it Phys. Rep.} {\bf 92} 1
\item[] Gottschalk B, Shlaer W J and Wang K H 1965 \PL {\bf 16} 294
\item[] Gottschalk B, Shlaer W J and Wang K H 1966 \NP {\bf 75} 549
\item[] Gottschalk B, Shlaer W J and Wang K H 1967 \NP {\bf A94} 491
\item[] Jones S, Madison D H, Franz A and Altick F 1993 \PR A {\bf 48} R22
\item[] K\"{o}v\'{e}r ´A and Laricchia G 1998 \PRL {\bf 80} 5309
\item[] Laricchia G, Armitage S, K¨ov´er A and Murtagh D J 2008 {\it Adv. At. Mol. Opt. Phy.} {\bf 56} 1
\item[] Lim K L and McCarthy I E 1966 \NP {\bf 88} 433
\item[] Macek J H, Sternberg J B, Ovchinnikov, Lee T G and Briggs J S 2009 \PRL {\bf 102} 143201
\item[] Macek J H 2010 {\it J. Phys.: Conf. Ser.} {\bf 212} 012008
\item[] Macek J H, Sternberg J B, Ovchinnikov S Yu and Briggs J S 2010 \PRL {\bf 104} 033201
\item[] Madelung E 1926 \ZP {\bf 40} 332
\item[] Malcolm G 2004 {\it Multidisciplinary Approaches To Visual Representations And Interpretations} (Amsterdam: Elsevier)
\item[] Marler J P and Surko C M 2005 \PR {\it A} {\bf 72} 062713
\item[] Meyer K R 1999 {\it Periodic Solutions of the N-Body Problem}, Lecture notes in mathematics vol 1719 (Springer-Verlag) p 25
\item[] Mueller D W, Lee C, Vermet C, Armitage S, Slaughter S, Hargrave L, Dorn A, Brunton J, Buckman S J and Sullivan J P 2012 {\it Bulletin of the American Physical Society, 43rd Annual Meeting of the APS Division of Atomic, Molecular and Optical Physics} {\bf 57} (5) 2012
\item[] Murray A J and Read F H 1993 \jpb {\bf 26} L359
\item[] Navarrete F, Della Picca R, Fiol J and Barrachina R O 2011 {\it Vortices in positron-atom ionization collisions}, XVI International Workshop on Low-Energy Positron and Positronium Physics (Maynooth, Ireland)
\item[] Ovchinnikov S Yu, Sternberg J B, Macek J H, Lee T-G and Schultz D R 2010 \PRL {\bf 105} 203005
\item[] Ovchinnikov S Y, Macek J H, Schmidt L Ph H and Schultz D R 2011 \PR A {\bf 83} 060701
\item[] Pauwels L 2006 {\it Visual Cultures of Science: Rethinking Representational Practices in Knowledge Building And Science Communication Interfaces} (Lebanon, NH: University Press of New England)
\item[] Surko C M, Gribakin G F and Buckman S J 2005 \jpb {\bf 38} R57
\item[] Thomas L H 1927 {\it Proc. R. Soc.} A {\bf 114} 561
\item[] Ullrich J, Moshammer R, D¨orner R, Jagutzki O, Mergel V, Schmidt-B¨ocking H and Spielberger L 1997 \JPB {\bf 30} 2917
\item[] Williams A I, K\"{o}v\'{e}r \'{A}, Murtagh D J and G. Laricchia G 2010  {\it J. Phys.: Conf. Ser.} {\bf 199} 012025
\endrefs

\end{document}